\input{aipcheck}

\documentclass[
    ,final            
  ]
  {aipproc}

\layoutstyle{6x9}

\begin{document}

\title{Constraints on single entity driven inflationary and radition eras}


\classification{98.80.Bp, 04.30.-w, 95.36.+x}
\keywords      {Inflationary era, Cosmological perturbations}

\author{Mariam Bouhmadi-L\'opez}{
  address={Centro Multidisciplinar de Astrof\'{\i}sica - CENTRA, Departamento de F\'{\i}sica, Instituto Superior T\'ecnico, Av. Rovisco Pais 1,1049-001 Lisboa, Portugal}
}

\author{Pisin Chen}{
  address={Department of Physics \& Graduate Institute of Astrophysics, National Taiwan University, Taipei 10617, Taiwan, R.O.C.},
altaddress={Leung Center for Cosmology and Particle Astrophysics,
National Taiwan University, Taipei 10617, Taiwan, R.O.C.}
}

\author{Yen-Wei Liu}{
  address={Department of Physics \& Graduate Institute of Astrophysics, National Taiwan University, Taipei 10617, Taiwan, R.O.C.}
  ,altaddress={Leung Center for Cosmology and Particle Astrophysics,
National Taiwan University, Taipei 10617, Taiwan, R.O.C.} 
}

\begin{abstract}
We present a model that attempts to fuse the inflationary era and the subsequent radiation dominated era under a unified framework so as to provide a smooth transition between the two. The model is based on a modification of the generalized Chaplygin gas.
We constrain the model observationally by mapping the primordial power spectrum of the scalar perturbations to the latest data of WMAP7. We compute as well the spectrum of the primordial gravitational waves as would be measured today. 
\end{abstract}

\maketitle


\section{Introduction}

While our knowledge about the universe has improved over the last decades with the advent of new observational data, there are several dark sides of the universe that have not been so far described from a fundamental point of view: what caused the initial inflationary era of the universe? what is the origin of dark matter? what is the fundamental cause of the current acceleration of the universe? Even though nowadays none of the previous issues have  a satisfactory answer, a parallel approach, that can shed some light on the dark sides of the universe, is a phenomenological one or a model building strategy.

The main goal of this paper is to obtain a phenomenologically consistent model for the early universe (inflationary and radiation dominated epochs) by properly modifying  the generalised Chaplygin gas (GCG) \cite{chaplygin}.
A first attempt in this direction has been recently carried out in \cite{BouhmadiLopez:2009hv} where a new scenario for the early universe was proposed. Such a scenario provides a  smooth transition between an early de Sitter-like phase and a subsequent radiation dominated era. Here, we give a more realistic model where the early inflationary phase of the universe is described by a ``quintessence'' inflationary phase \cite{BouhmadiLopez:2011kw}. This phase will be connected to a radiation dominated phase at later time \cite{BouhmadiLopez:2011kw}. The model can be described through a scalar field or a Chaplygin gas inspired model. We will then analyze the possible imprints of such a gas in the primordial power spectrum of scalar perturbations and the power spectrum of the stochastic background of gravitational waves. 
\section{The model}

We start considering a  model corresponding to an inflationary period with a  ``quintessence'' like behavior (described by a power law expansion) and  followed by a radiation dominated epoch. The matter content of the universe can then be modeled \textit{\`a la} Chaplygin gas as
\begin{equation}
\rho=\left(\frac{A}{a^{1+\beta}}+\frac{B}{a^{4(1+\alpha)}}\right)^{1/(1+\alpha)},
\label{rho}
\end{equation}
where $A,B,\alpha,\beta$ are constants such that  $2(1+\alpha)<1+\beta<0$.
Such a model generalize the scenario presented in \cite{BouhmadiLopez:2009hv}. As this matter content is not interacting with any other fluid it fulfills the following equation of state:
\begin{equation}
p=\frac13\rho+\frac{A}{3(1+\alpha)}\frac{1+\beta-4(1+\alpha)}{a^{1+\beta}} \rho^{-\alpha}.
\label{p2}
\end{equation}
This equation shows clearly that we recover the model discussed by one of us in \cite{BouhmadiLopez:2009hv} for $\beta\rightarrow - 1$. We would like to highlight that the equation of state (\ref{p2}) has been previously analysed in \cite{Chimento:2009sh}.

\begin{figure}[t]
  \includegraphics[width=6cm]{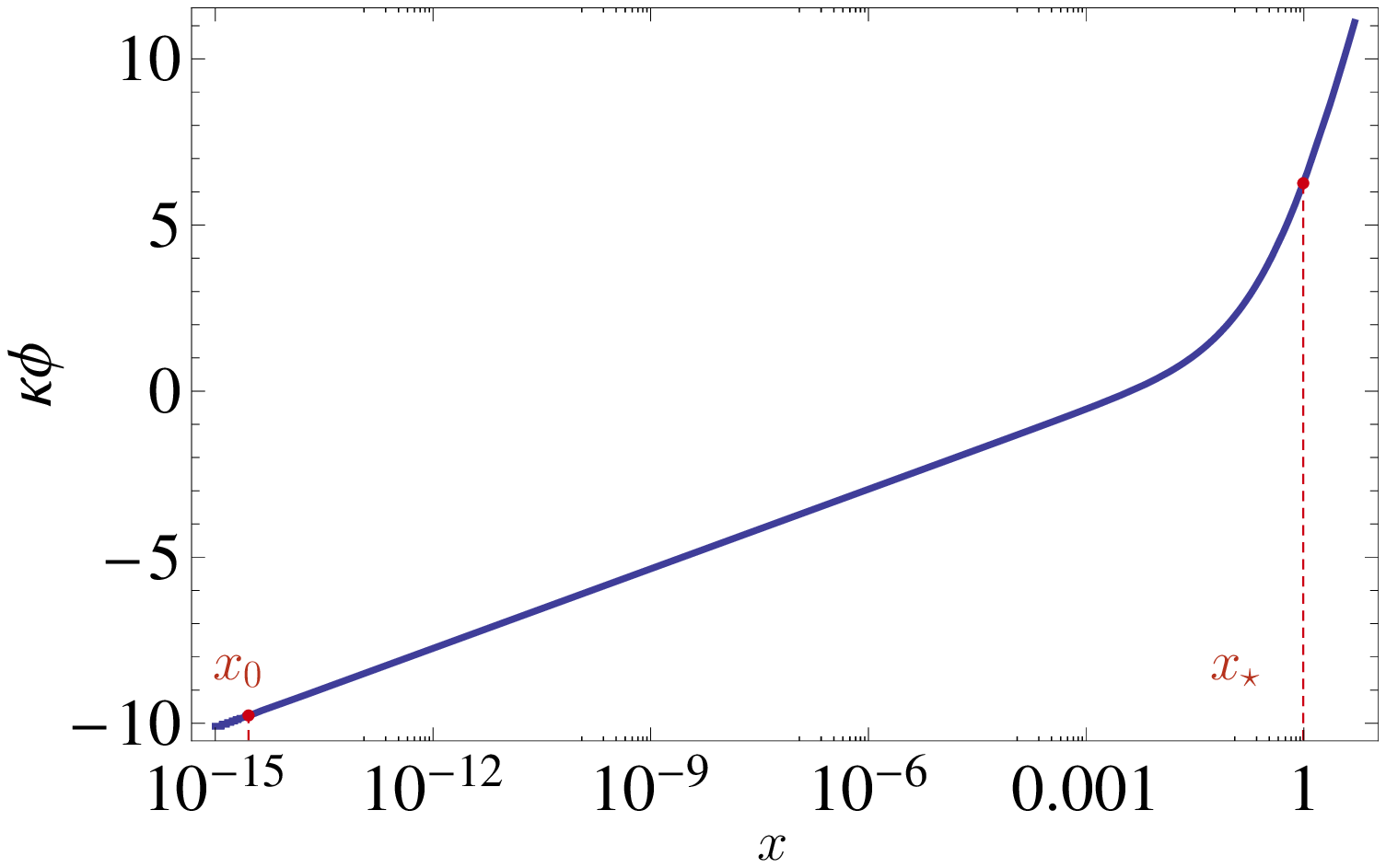}\hspace*{1.5cm}\includegraphics[width=6cm]{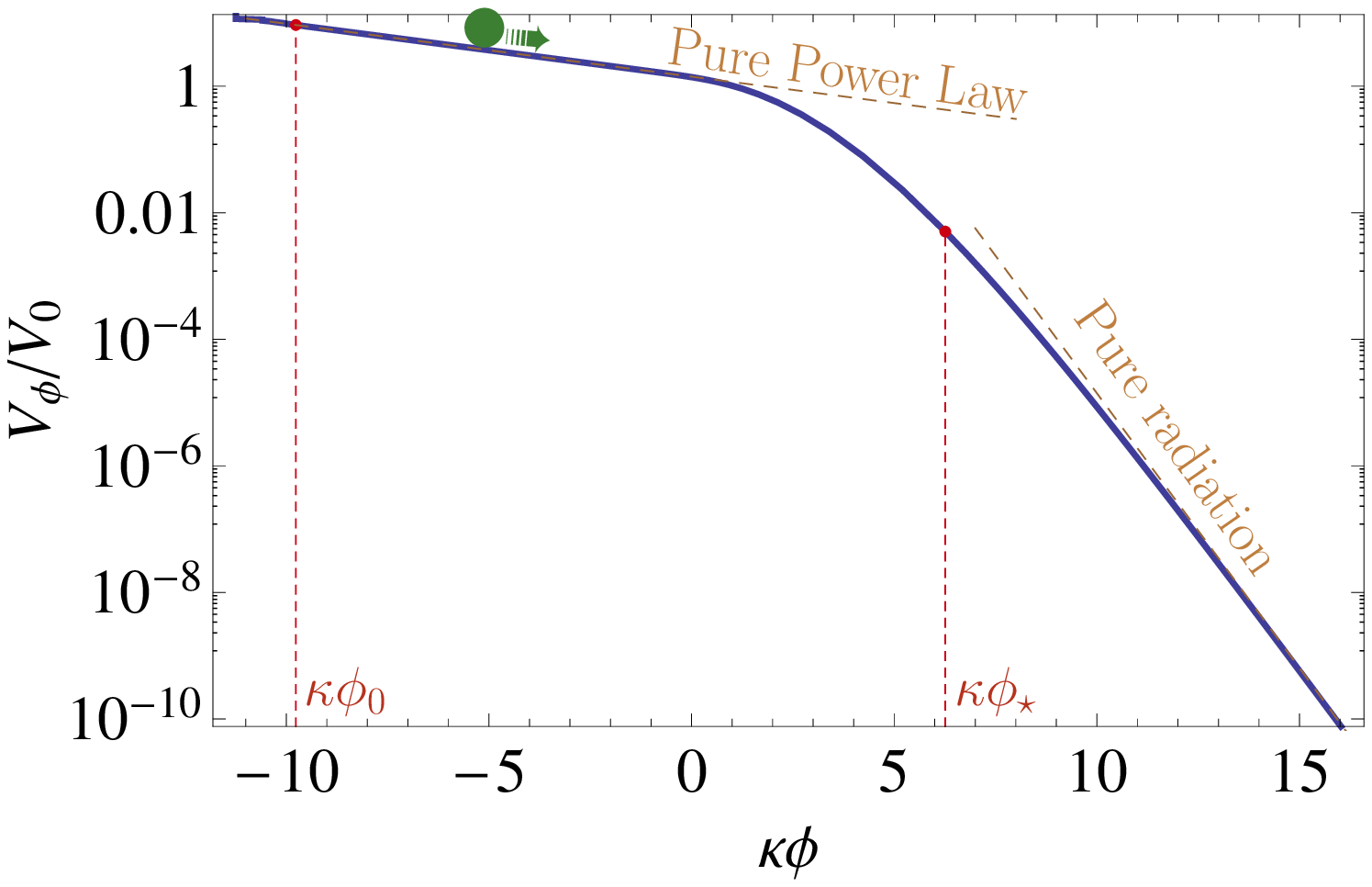}
	\caption{The left hand side (lhs) curve corresponds to $\phi$ against $x=(B/A)a^q$ where $q=1+\beta-4(1+\alpha)$.
The rhs curve corresponds to  $V(\phi)$ against $\phi$, where $V_0=A^{1/(1+\alpha)}\left(A/B\right)^{-(1+\beta)/[q(1+\alpha)]}$. The values $\phi_0,\,x_0$ and $\phi_\star,\,x_\star$ correspond to the moments when the pivot scale $k_0=0.002\, \textrm{Mpc}^{-1}$ exists the horizon and the end of inflation, respectively.}
	\label{Fig1}
\end{figure}

The inflationary dynamics of the model presented in Eq.~(\ref{rho}) can be  described through a minimally coupled scalar field, $\phi$, with a potential, $V(\phi)$, whose shape is shown in Fig.~\ref{Fig1} (please see Ref.~\cite{BouhmadiLopez:2011kw} for more details). 

\section{Cosmological imprints}

Inflation generates density perturbations that seeds the structure of the present universe. Those density perturbations have been constrained through observations of the cosmic microwave background (CMB). Next, we will constrain the model introduced before by using the measurements of  WMAP7 \cite{Komatsu:2010fb} for the  power spectrum of the comoving curvature perturbations, $P_s=2.45\times 10^{-9}$, and its index, $n_s=0.963$. These measurements correspond to a pivot scale $k_0=0.002\, {\mathrm{Mpc^{-1}}}$ \cite{Komatsu:2010fb}.

\begin{figure}[t]
\includegraphics[width=6cm]{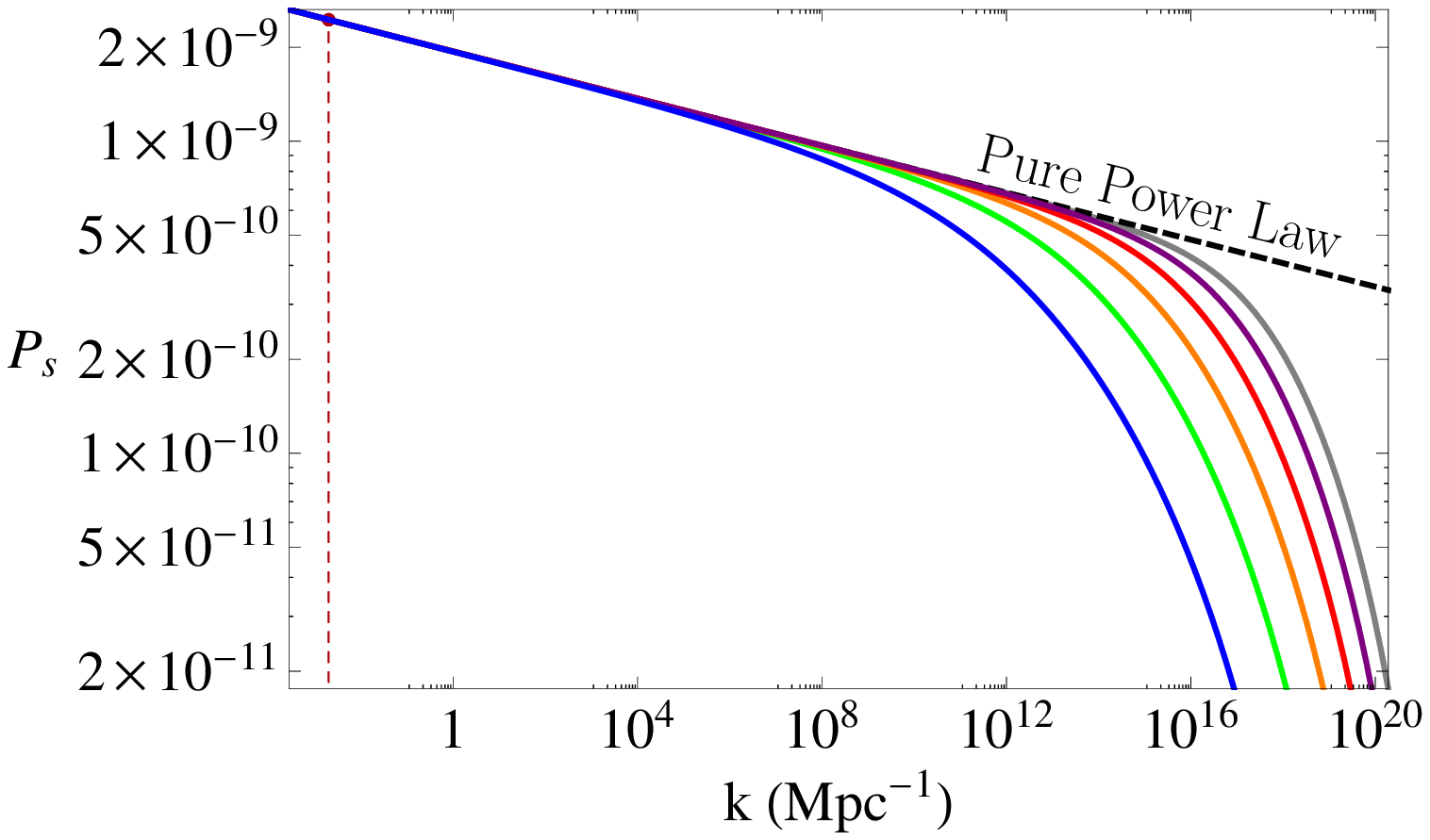}\hspace{1.5cm}\includegraphics[width=6cm]{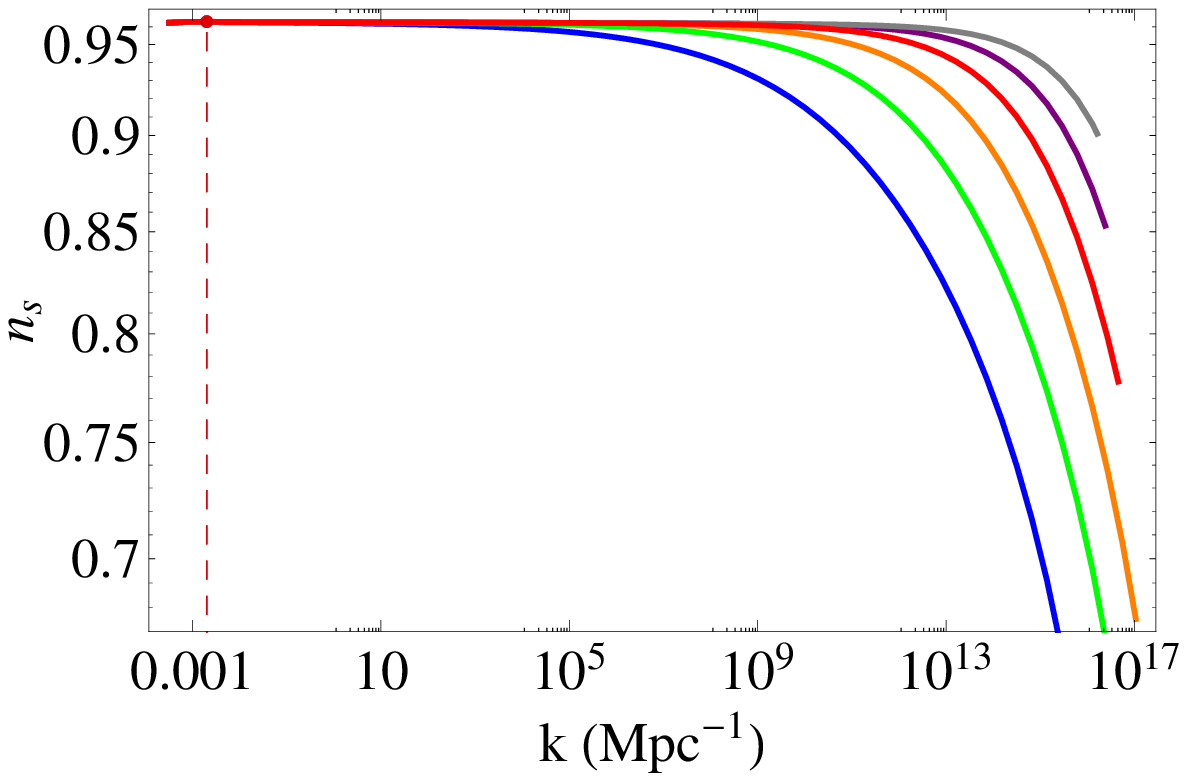}
\caption{Primordial power spectrum $P_s(k)|_{k=aH}$ and the spectral index $n_s$ against $k$ for six different values of $\alpha$. The dashed black line is the pure power law inflation, and the vertical dashed red line locates the pivot scale $k_0=0.002\textrm{Mpc}^{-1}$. We can see that all these lines merge when small $k$; i.e. large scale, exits the horizon. The grey, violet, red, orange, green and blue curve correspond respectively to $\alpha=-1.1,-1.09,-1.08,-1.07,-1.06,-1.05.$}
\label{Fig2}
\end{figure}

The comoving curvature perturbations is determined by the fluctuations of the scalar field $\phi$. The corresponding power spectrum for the field $\nu_k$ is \cite{Langlois:2010xc}
\begin{equation}
2\pi^2k^{-3}P_s(k)=\frac{|\nu_k|^2}{z^2}\label{powerspectrum},
\end{equation}
where $z=\frac{a}{H}\frac{d\phi}{dt}$ and $t$ is the cosmic time. The field $\nu_k$ satisfies in the Fourier space  the equation \cite{Langlois:2010xc}
\begin{equation}
\frac{d^2\nu_k}{d\eta^2}+\left(k^2-\frac{1}{z}\frac{d^2z}{d\eta^2}\right)\nu_k=0,\quad \eta=\textrm{conformal time}
\label{seof}.
\end{equation}
We show in Fig.~\ref{Fig2} our results for the power spectrum of the comoving curvature perturbations and its index where we have fixed the values of the parameters of the model as follows: (i) $B$ is fixed by the current amount of radiation in the universe, (ii) for a given parameter $\alpha$, the parameter $\beta$ is fully determined by the measurement of $n_s$ and (iii) the parameter $A$ is fixed such that $P_s=2.45\times 10^{-9}$ at the pivot scale $k_0=0.002\, {\mathrm{Mpc^{-1}}}$. Finally, we have imposed that when the wavelength of a given mode $k$ is much smaller than the Hubble radius $k\gg aH$, the effect of curvature can be neglected on $\nu_k$ and, therefore, the result reduces to that of a  flat Minkowski spacetime (when $k\gg aH$) \cite{Langlois:2010xc}. 

\begin{figure}[t]
\centering
\includegraphics[width=6cm]{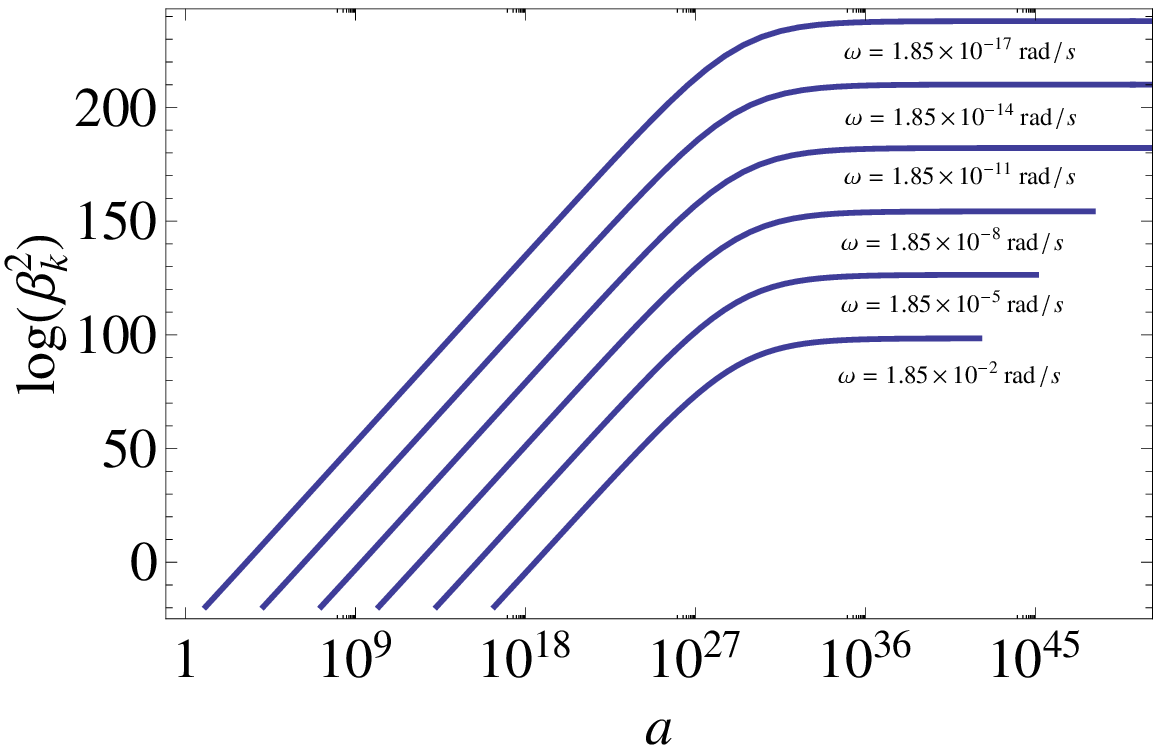}\hspace*{1.5cm}\includegraphics[width=6cm]{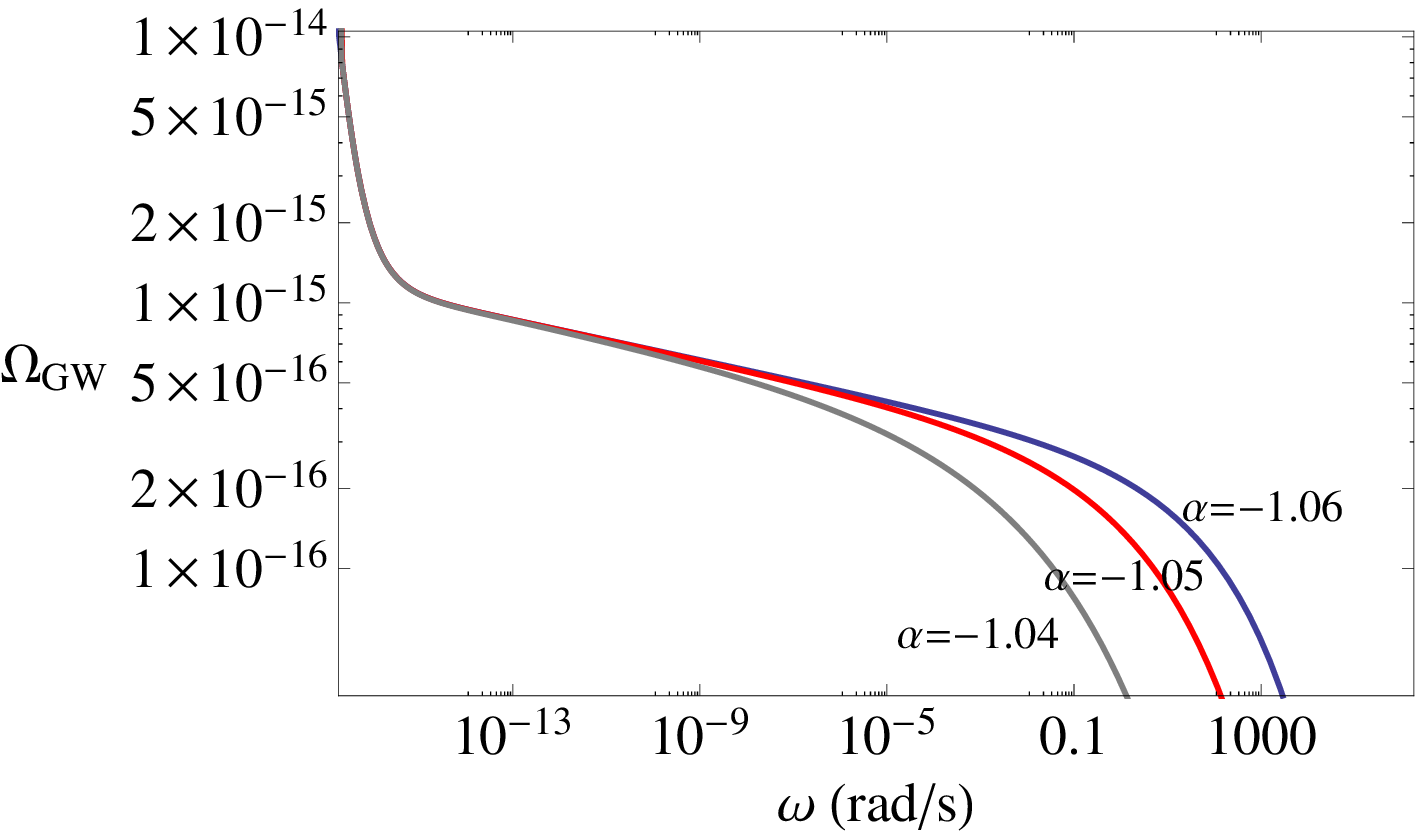}
\caption{On the lhs figure, we show examples of $|\beta_k|^2=N_k$ for different frequencies. On the rhs figure we show  $\Omega_{GW}$ against  $\omega$ for different value of $\alpha$ in this GCG model: the blue line refers to $\alpha=-1.06$, the red one refers to $\alpha=-1.05$, and the grey one to $\alpha=-1.04$.}
\label{Fig3}
\end{figure}

Similiraly, we can obtain the spectrum of the gravitational waves (GWs) using the method of the Bogoliubov coefficients \cite{Parker:1969au}. In particular, one of these coefficients, which we will denote $\beta_k$, gives the number of gravitons $N_k$,  $N_k=|\beta_k|^2$, created as the universe expands. In fact, the dimensionless relative logarithmic energy spectrum of the gravitational waves, $\Omega_{\mathrm{GW}}$, at the present time reads \cite{Sa:2008yq}:
\begin{equation}
\Omega_{GW}(\omega,\tau_0)\equiv\frac{1}{\rho_c(\tau_0)}\frac{d\rho_{GW}}{d\ln\omega}(\tau_0)=\frac{\hbar\kappa^2}{3\pi^2 c^5 H^2(\tau_0)}\omega^4\beta_k^2(\tau_0).
\label{spectrum}\end{equation}
The parameter $\rho_{\rm GW}$ is the energy density of GWs and  $\omega$  the respective
 angular frequency; $\rho _{\rm c}$ and $H$ are the critical density of the
universe and Hubble parameter, respectively, evaluated at the present time. Our results are shown in Fig.~\ref{Fig3}.

\section{Conclusions}

We propose a phenomenological model for the early universe where there is a
smooth transition between an early ``quintessence'' phase and a radiation dominated era.
We constrain the model observationally by mapping the primordial power spectrum of the scalar perturbations to the latest data of WMAP7. We compute as well the spectrum of the primordial gravitational waves as would be measured today.


\begin{theacknowledgments}
M. B. L. is supported by the Portuguese Agency Fundac\~ao para a Ci\^{e}ncia e Tecnologia through SFRH/BPD/26542/2006 and PTDC/ 
FIS/111032/2009. P.C. and Y.W.L. are supported by Taiwan National Science Council under Project No. NSC 97-2112-M-002-026-MY3 and by Taiwan's National Center for Theoretical Sciences. P.C. is in addition supported by US Department of Energy under Contract No. DE-AC03-76SF00515.

\end{theacknowledgments}



\bibliographystyle{aipproc}   

\bibliography{sample}



\end{document}